\documentclass[prl,twocolumn,superscriptaddress,showpacs]{revtex4}
\usepackage{amsthm,amsmath,amssymb}
\usepackage[utf8x]{inputenc}
\usepackage[dvipdfm]{graphicx}
\usepackage[english]{babel}

\newtheorem{theorem}{Theorem}

\begin{document}
\title{All scale-free networks are sparse}
\author{Charo I. \surname{Del Genio}}
	\affiliation{Max-Planck-Institut f\"ur Physik komplexer Systeme, N\"othnitzer Stra{\ss}e 38, 01187 Dresden, Germany}
\author{Thilo \surname{Gross}}
	\affiliation{Max-Planck-Institut f\"ur Physik komplexer Systeme, N\"othnitzer Stra{\ss}e 38, 01187 Dresden, Germany}
\author{Kevin E. \surname{Bassler}}
	\affiliation{\mbox{Department of Physics, 617 Science and Research 1, University of Houston, Houston, Texas 77204-5005, USA}}
	\affiliation{\mbox{Texas Center for Superconductivity, 202 Houston Science Center, University of Houston, Houston, Texas 77204-5002, USA}}

\date{\today}

\begin{abstract}
We study the realizability of scale free-networks with a given degree sequence,
showing that the fraction of realizable sequences undergoes two first-order transitions
at the values 0 and 2 of the power-law exponent. We substantiate this finding by
analytical reasoning and by a numerical method, proposed here, based on extreme
value arguments, which can be applied to any given degree distribution. Our results
reveal a fundamental reason why large scale-free networks without constraints on
minimum and maximum degree must be sparse.
\end{abstract}


\maketitle

Many complex systems can be modeled as networks, i.e., as a set of connections
(edges) linking discrete elements (nodes)~\cite{Alb02,New03,Boc06}. A characteristic
of a network that affects many physical properties is its degree distribution
$P\left(k\right)$, the probability of finding a node with $k$ edges. Considerable
attention has been paid to scale-free networks, in which the degree distribution
follows a power-law, $P\left(k\right)\sim k^{-\gamma}$~\cite{Cal07,Pri65,Red98,AlbXX,AmaXX,New01,Vaz02}.
In particular, scale-freeness has been shown to have important implications in
the thermodynamic limit. For studying
the properties of scale-free networks, several generative models have been proposed~\cite{Alb02,New03,Boc06,Cal07}.
However, no models creating networks with $\gamma<2$ have been found~\cite{Dor00,Kra00},
and $\gamma<2$ is observed only in networks that are relatively small or in which
the power-law behavior has some cutoff~\cite{New01}. In this Letter, we explain
the absence of large networks that exhibit a power-law with $0<\gamma<2$ in the
tail of the distribution. Specifically, we show that fundamental constraints exist
that prevent the realization of any such network.

It is well known that the mean degree of scale-free distributions
with exponents $\gamma$ less than 2 diverges in the thermodynamic
limit, i.e., when the number of nodes $N\rightarrow\infty$~\cite{New03}.
Scale-free networks with $\gamma<2$ would therefore be called \emph{dense}
networks, whereas networks $\gamma>2$ are \emph{sparse}. While sparseness
is a common property, which is regularly exploited in data storage
and algorithms, also many examples of dense networks are known~\cite{Spi03,Ton04,Hag08}.
It is thus reasonable to ask why there are no examples of dense scale-free
networks. We answer this question by showing that dense networks
with a power-law degree distribution must have $\gamma<0$. Calling
such networks scale-free is at best dubious because they would not
exhibit the characteristic properties commonly associated with scale-freeness
for $N\rightarrow\infty$.

The absence of dense scale-free networks is explained by a discontinuous
transition in the realizability of such networks. Below, we show numerically,
analytically, and by a hybrid method proposed here, that the probability
of finding a scale free-network with a given $\gamma$ is 0 for $0<\gamma<2$.
We emphasize that these results are not contingent on a specific generative
model, but arise directly from fundamental mathematical constraints.

The generation of scale-free networks with a given degree distribution
can be considered as a two-step procedure. First, one creates a number
of nodes and assigns to each node a number of connection ``stubs'' drawn
from the degree distribution. The realization of the degree distribution
that is thus created is called \emph{degree sequence}. Second, one connects
the stubs such that every stub on a given node links to a stub on a different
node, without forming self-loops or double links. However, not every degree
sequence can be realized in a network. Sequences that admit realizations
as simple graphs are called graphical, and their realizability property
is commonly referred to as graphicality~\cite{Erd60}. Graphicality fails
trivially if the number of stubs is odd, as one needs two stubs to form
every link, or if the degree of any node is equal to or greater than the
number of nodes, as it would be impossible to connect all its stubs to
different nodes. Below we do not consider sequences for which graphicality
is such trivially violated, but note that further conditions must be met
for a sequence to be graphical~\cite{Kim09,Del10}.

The main result used for testing the graphicality of a degree sequence
is the Erdős-Gallai theorem, stated here as reformulated
in~\cite{Del10} using recurrence relations:
\begin{theorem}\label{EGtest}
Let $\mathcal D=\left\lbrace d_0, d_1, \dotsc, d_{N-1}\right\rbrace$ be a non-increasing
degree sequence on $N$ nodes. Define $x_k = \min\left\lbrace i : d_i<k+1\right\rbrace$
and $k^\star = \min\left\lbrace i : x_i<i+1\right\rbrace$. Then, $\mathcal D$ is graphical
if and only if $\sum_{i=0}^{N-1}d_i$ is even, and
\begin{equation}\label{EGin}
 L_k\leqslant R_k\quad\forall\:0\leqslant k<N-1\:,
\end{equation} 
where $L_k$ and $R_k$ are given by the recurrence relations
\begin{align}
L_0 &= d_0\\
L_k &= L_{k-1} + d_k
\end{align}
and
\begin{align}
R_0 &= N-1\\
R_k &= \left\lbrace\begin{array}{ll}R_{k-1}+x_k-1 & \forall k<k^\star\\R_{k-1}+2k-d_k & \forall k\geqslant k^\star\end{array}\right.\label{rhsr}\:.
\end{align}
\end{theorem}
This formulation of the theorem has the advantage over the traditional one~\cite{Erd60}
of allowing a very fast implementation of a graphicality test~\cite{Del10}.

Equivalently, graphicality can be tested by a recursive application of the Havel-Hakimi
theorem~\cite{Hav55,Hak62}:
\begin{theorem}
 A non-increasing degree sequence $\mathcal D=\left\lbrace d_0, d_1, \dotsc, d_{N-1}\right\rbrace$
is graphical if and only if the sequence $\mathcal D'=\left\lbrace d_1-1, d_2-1, \dotsc, d_{d_0}-1, \dotsc, d_{N-1}\right\rbrace$
is graphical.
\end{theorem}

\begin{figure}
\centering
\vspace*{-10pt}
{\includegraphics[width=0.45\textwidth]{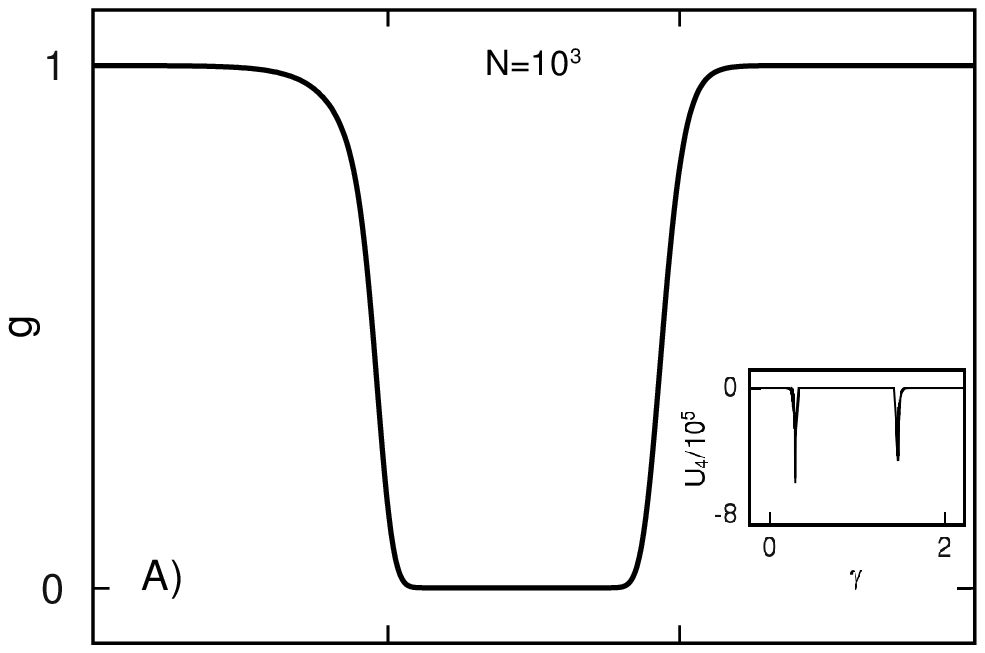}
\vspace*{-26pt}}
{\includegraphics[width=0.45\textwidth]{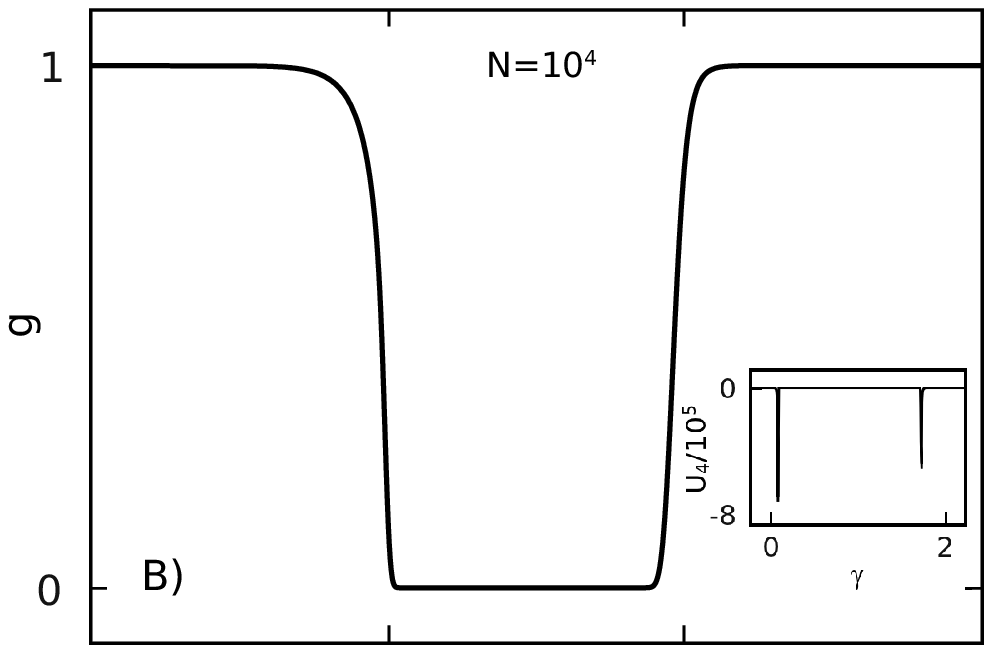}
\vspace*{-26pt}}
{\includegraphics[width=0.45\textwidth]{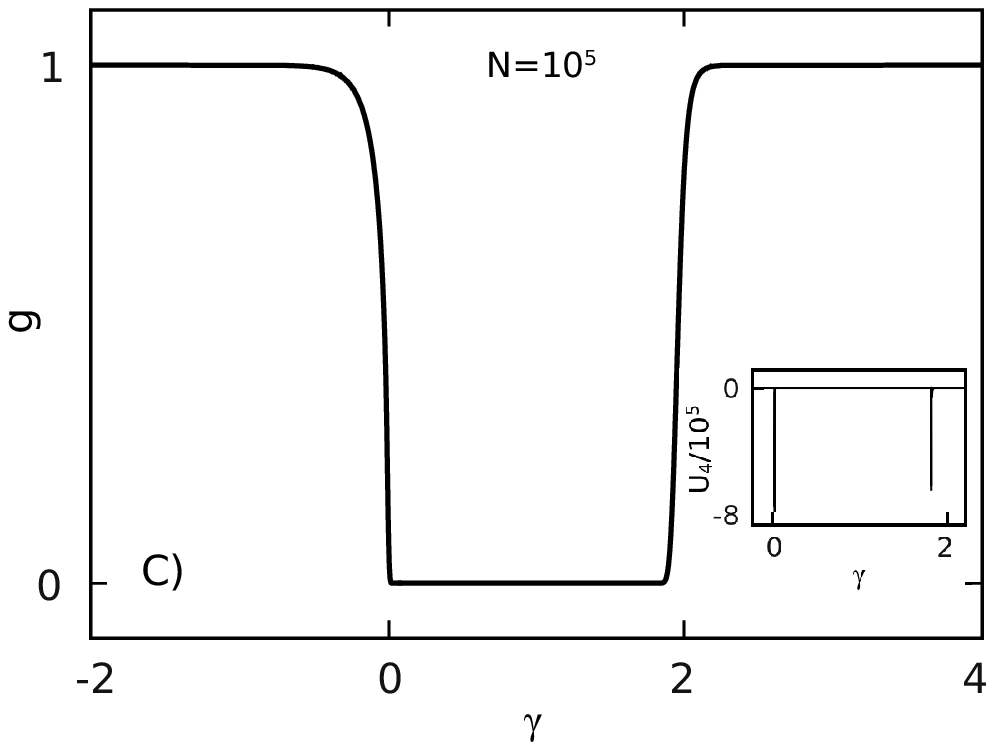}
\vspace*{-15pt}}
\caption{\label{Fig1}Graphicality transitions in scale-free networks.
The plots of graphical fraction $g$ vs.\ exponent $\gamma$
show transitions at $\gamma=0$ and $\gamma=2$. Binder's cumulants,
in the insets, identify the character of the transitions
and the transition points.}
\end{figure}

To investigate the dependence of the graphicality of scale-free networks
on the power-law exponent $\gamma$, we performed extensive numerics, generating
ensembles of sequences of random power-law distributed integers with range
between 1 and $N-1$ and $\gamma$ between $−2$ and 4. We tested each
sequence for graphicality by applying Th.~\ref{EGtest}, and computed for
each $\gamma$ the graphical fraction
\begin{equation*}
 g = \frac{G}{E}\:,
\end{equation*}
where $G$ is the total number of graphical sequences in the ensemble
and $E$ is the number of sequences in the ensemble with an even degree
sum. The results, plotted in Fig.~\ref{Fig1}, clearly show two graphicality
transitions: For very large and very small exponents almost all sequences
are graphical. However, at intermediate exponents there is a pronounced
gap where almost no sequence is graphical. The transitions between the
two phases become steeper and the transition points approach $\gamma=0$,
and $\gamma=2$ as the system size is increased.

The dependence of $g$ on sequence length strongly suggests that both
transitions are first-order. To verify their character, we studied Binder's
cumulants $U_4 \equiv 1-\left\langle g^4\right\rangle/{\left\langle g^2\right\rangle}^2$~\cite{Bin81,Bin81_2}.
For continuous transitions, the cumulants for different system sizes
lie within a finite interval and cross at the critical point, whereas
for first-order transitions the curves are flat, except for a diverging
negative minimum whose position converges to the transition point with
increasing system size~\cite{Bin84,Vol93,Lan00}. In the present system
the cumulants confirm that the graphicality transitions at $\gamma=0$
and $\gamma=2$ are first order.

To understand the origin of the transitions, we focus on the scaling of the
largest degrees and of the number of lowest degree nodes in the sequences.
Below, we show that the first two largest degrees are of order $O\left(N\right)$
for $\gamma\leqslant2$, while they grow sublinearly with $N$ for $\gamma>2$.
Also, the number of nodes with degree of order $O\left(1\right)$ increases
linearly with $N$ for $\gamma\geqslant0$ and decreases like $N^\gamma$ for
$\gamma<0$. Then, the transitions can be understood as follows: If we tried
to construct a scale-free network with $\gamma$ between 0 and 2, following
the Havel-Hakimi algorithm, $O\left(N\right)$ nodes with unitary degree would
be used to place the connections involving the first node, and then there would
be no way to place all the needed edges involving the node with the second
largest degree. Conversely, when $\gamma<0$, all but a vanishingly small fraction
of nodes have a degree of order $O\left(N\right)$, and for $N\rightarrow\infty$
all the nodes are able to form as many connections as needed.

To see this, calculate the expected maximum degree of a scale-free sequence
\begin{equation}\label{disc}
 \widehat d = \max\left\lbrace x:\ N\sum_{k=x}^{N-1}\frac{k^{-\gamma}}{H_{N-1,\gamma}}\geqslant1\right\rbrace\:,
\end{equation}
where $H_{a,b}$ is the $a^\mathrm{th}$ generalized harmonic number of exponent $b$
\begin{equation*}
 H_{a,b} = \sum_{t=1}^at^{-b}\:.
\end{equation*}
When $N\gg1$, Eq.~\ref{disc} becomes
\begin{equation}\label{inteq}
 N\int_x^{N-1}\frac{k^{-\gamma}}{H_{N-1,\gamma}}\mathrm{dk}=1\:.
\end{equation}
Because of the dependence of the behavior of the generalized harmonic number
on the exponent, we solve this equation for different values of $\gamma$.

Solving the integral for $\gamma>1$ gives
\begin{equation}\label{intsol}
 \frac{N}{\left(1-\gamma\right)H_{N-1,\gamma}}\left[\left(N-1\right)^{1-\gamma}-x^{1-\gamma}\right]=1\:.
\end{equation}
Equation~\ref{intsol}, for $N\gg1$ implies
\begin{equation*}
 x=\left[\frac{N}{\left(\gamma-1\right)H_{N-1,\gamma}}\right]^{\frac{1}{\gamma-1}}\sim N^{\frac{1}{\gamma-1}}\:.
\end{equation*}
Because of the upper bound of the degrees of a sequence at $N-1$,
if $1<\gamma\leqslant2$, the value of the largest degree grows linearly with
the number of nodes $N$.

For $\gamma=1$, the integral in Eq.~\ref{inteq} gives
\begin{equation*}
 \log\left(N-1\right)-\log\left(x\right)=\frac{H_{N-1}}{N}\:,
\end{equation*}
where $H_{N-1}=H_{N-1,1}$ is the $\left(N-1\right)^\mathrm{th}$
harmonic number. To solve the above equation note that the right-hand
side vanishes in the limit of large $N$. This can be seen by an
application of l'Hôpital's rule, noticing that for $\gamma\geqslant0$
\begin{equation*}
 \frac{\partial}{\partial \mathrm{N}} H_{N-1,\gamma}=\gamma\left[\zeta\left(\gamma+1\right)-H_{N-1,\gamma+1}\right]
\end{equation*}
and
\begin{equation*}
 \lim_{N\rightarrow\infty}H_{N,\gamma}=\zeta\left(\gamma\right)\:,
\end{equation*}
where $\zeta$ is Riemann's zeta function.
Then, the solution of the equation in the thermodynamic limit
is $x\sim N$.

Next, for $0\leqslant\gamma<1$, Eq.~\ref{inteq} yields Eq.~\ref{intsol}, hence
\begin{equation}\label{normsol}
 \left(N-1\right)^{1-\gamma}-x^{1-\gamma}=\frac{\left(1-\gamma\right)H_{N-1,\gamma}}{N}\:.
\end{equation}
As in the previous case, the right hand side vanishes
for large $N$, and the solution is that $x\sim N$.

Finally, for $\gamma<0$, from Eq.~\ref{intsol} one gets
again Eq.~\ref{normsol}. However, in this case the right-hand
side grows as $N^{-\gamma}$. Since $\gamma$ is negative,
one can rewrite Eq.~\ref{normsol} for large $N$ as
\begin{equation*}
 \left(N-1\right)^{1+\left|\gamma\right|}-x^{1+\left|\gamma\right|}=N^{\left|\gamma\right|}\:,
\end{equation*}
which implies again that $x\sim N$.

The same arguments can be applied to the scaling of the second largest
degree, with identical results. Now, consider the number $A$ of nodes
with unitary degree. For large $N$, $A=N/H_{N-1,\gamma}$. Thus, when $\gamma\geqslant0$,
$A\sim N$, whereas, if $\gamma<0$, then $A\sim N^\gamma$.

Then, to formally check the transition mechanism, explicitly
write inequality~\ref{EGin} for $k=1$. The left-hand-side consists
of the sum of the largest and the second largest degrees, which
can be obtained using the same argument as above. To compute
the right hand side, first notice that $k^\star\geqslant 2$.
In fact, by definition it cannot be 0, as this would
imply that the highest degree in the sequence would be 0. Moreover,
in our case it cannot be 1, as this would imply that the second
highest degree in the sequence would be 1, in contradiction
with what demonstrated above. Also, by the definition of $x_k$,
it follows that $x_1=N-A$. Thus, applying Eq.~\ref{rhsr},
the right hand side is simply $2N-2-A$. Therefore, the inequality reads
\begin{widetext}
\begin{equation}
 \left[\frac{\left(\gamma-1\right)H_{N-1,\gamma}}{N}+\left(N-1\right)^{1-\gamma}\right]^{\frac{1}{1-\gamma}}+\left[\frac{2\left(\gamma-1\right)H_{N-1,\gamma}}{N}+\left(N-1\right)^{1-\gamma}\right]^{\frac{1}{1-\gamma}}\leqslant2N-2-\frac{N}{H_{N-1,\gamma}}\:.
\end{equation}
\end{widetext}
A numerical solution shows that for $N\gg1$ the above inequality
is indeed satisfied only when $\gamma<0$ or $\gamma>2$, confirming
the transition mechanism. Notice, however, that when $N\rightarrow\infty$
almost all the nodes are fully connected for $\gamma<0$, and
thus it is not appropriate to refer to such networks as scale-free.

One can also study the inequality in the presence
of a cutoff in the distribution, by replacing every instance
of the natural upper bound on the degrees, $\left(N-1\right)$,
with the cutoff value. Cutoffs have been observed in real-world
networks~\cite{AmaXX,New01}, and are sometimes imposed
for different purposes, such as
making the degree-degree correlations uniform~\cite{Mos02,Cat05}.
Notably, their effect is making the inequality always satisfied,
and the transitions disappear.

The above treatment indicates that, by applying extreme value arguments,
one can build a finite length sequence $S=\left\lbrace s_0, s_1, \dotsc, s_{N-1}\right\rbrace$
for the purpose of studying the graphicality of infinite systems. A finite
sequence maximizing the degrees of the nodes for a given degree distribution
will best approximate the graphicality of an infinite sequence, especially
since broken graphicality is always caused by an excess of stubs in some
subset of nodes. Therefore, for a length $N$, and any degree distribution
$P\left(d\right)$, the elements of the sequence are given by the family
of functionals
\begin{equation}\label{degmax}
 s_i = \max\left\lbrace s^\star : N\sum_{d=s^\star}^{d_M}P\left(d\right)\geqslant i+1\right\rbrace\:,
\end{equation}
where $d_M$ is the largest allowed degree. In general, $d_M=N-1$,
but the full generality of its value allows cutoffs to be accounted
for. Increasing the number of nodes in the representative sequence
will improve the accuracy in the determination of transition points,
as it will better approximate an infinitely large system.

\begin{figure}
 \centering
\vspace*{-7pt}
{\includegraphics[width=0.45\textwidth]{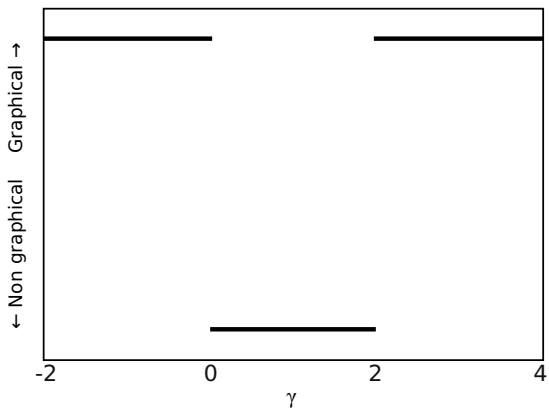}
\vspace*{-15pt}}
\caption{\label{Fig2}Graphicality of the degree maximizing sequence
($N=10^6$) given by Eq.~\ref{degmax} for scale-free distributions
vs.\ exponent $\gamma$. The graphicality transitions points are
correctly identified.}
\end{figure}

We computed degree maximizing sequences for power-law
distributions, and tested them for graphicality. The results, shown
in Fig.~\ref{Fig2}, are consistent with the simulations and with
the analytical treatment, showing once again transitions at $\gamma=0$
and $\gamma=2$.

In conclusion, we showed that the graphicality of power-law degree sequences
undergoes two discontinuous transitions at the values~0 and~2 of the exponent
$\gamma$. In the limit of a large number of nodes, no network with an unbounded
power-law degree distribution with $0<\gamma<2$ can exist. We emphasize that
this result arises directly from mathematical constraints on the degrees of
the nodes, and is thus independent
of the specific procedure used for generating the network. It explains why the
scale-free networks commonly observed in nature have $\gamma>2$ or have a cutoff.
Established procedures may yield $0<\gamma<2$
when applied to data from a given finite network. However, when the network
grows or more data is acquired either a cutoff must exist or $\gamma$ must increase
above 2 (or decrease below 0). It is possible to generate large and dense networks
with a power-law degree distribution with $\gamma<0$, but these networks should
not be denoted as scale-free as they do not exhibit the properties that are
commonly associated with scale-freeness. Any large scale-free network is thus
sparse, either because $\gamma>2$ or because of the presence of a cutoff. This
insight is reassuring as it implies that also numerical methods which are often
needed for analyzing scale free networks will continue to scale favorably with
increasing network size. 

\begin{acknowledgments}
The authors gratefully acknowledge Zoltán Toroczkai, Hyunju Kim
and Chiu-Fan Lee for fruitful discussions and helpful comments
on the manuscript. KEB was supported by NSF grant No.~DMR-0908286.
\end{acknowledgments}


\begin{thebibliography}{99}
\bibitem{Alb02}   R. Albert and A.-L. Barabási, Rev.~Mod.~Phys. \textbf{74}, 47 (2002).
\bibitem{New03}   M.~E.~J. Newman, SIAM~Review \textbf{45}, 167 (2003).
\bibitem{Boc06}   S. Boccaletti et al.\ Phys. Rep. \textbf{424}, 175 (2006).
\bibitem{Cal07}   G. Caldarelli, \textit{Scale-free networks -- complex webs in nature and technology} (Oxford University Press, Oxford, United Kingdom, 2007).
\bibitem{Pri65}   D.~J. de~Solla Price, Science \textbf{149}, 510 (1965).
\bibitem{Red98}   S. Redner, Eur.~J.~Phys.~B \textbf{4}, 131 (1998).
\bibitem{AlbXX}   R. Albert, H. Jeong, and A.-L. Barabási, Nature \textbf{401}, 130 (1999); H. Jeong et al.\ Nature \textbf{407}, 651 (2000); H. Jeong et al.\ Nature \textbf{411}, 41 (2001).
\bibitem{AmaXX}   L.~A.~N. Amaral et al.\ Proc.~Natl.~Acad.~Sci.~USA \textbf{97}, 11149 (2000); F. Liljeros et al.\ Nature \textbf{411}, 907 (2001).
\bibitem{New01}   M.~E.~J. Newman, Proc.~Natl.~Acad.~Sci.~USA \textbf{98}, 404 (2001).
\bibitem{Vaz02}   A. Vázquez, R. Pastor-Satorras, and A. Vespignani, Phys.~Rev.~E \textbf{65}, 066130 (2002).
\bibitem{Dor00}   S.~N. Dorogovtsev, J.~F.~F. Mendes, and A.~N. Samukhin, Phys.~Rev.~Lett. \textbf{85}, 4633 (2000).
\bibitem{Kra00}   P.~L. Krapivsky, S. Redner, and F. Leyvraz, Phys.~Rev.~Lett. \textbf{85}, 4629 (2000).
\bibitem{Spi03}   V. Spirin and L.~A. Mirny, Proc.~Natl.~Acad.~Sci.~USA \textbf{100}, 12123 (2003).
\bibitem{Ton04}   A.~H.~Y. Ton et al.\ Science \textbf{303}, 808 (2004).
\bibitem{Hag08}   P. Hagmann et al.\ PLoS~Biology \textbf{6}, 1479 (2008).
\bibitem{Kim09}   H. Kim et al.\ J.~Phys.~A -- Math.~Theor. \textbf{42}, 392001 (2009).
\bibitem{Del10}   C.~I. Del~Genio et al.\ PLoS~One \textbf{5}, e10012 (2010).
\bibitem{Erd60}   P. Erdős and T. Gallai, Mat.~Lapok \textbf{11}, 477 (1960).
\bibitem{Hav55}   V. Havel, Časopis Pěst Mat \textbf{80}, 477 (1955).
\bibitem{Hak62}   S.~L. Hakimi, J.~Soc.~Ind.~Appl.~Math.\ \textbf{10}, 496 (1962).
\bibitem{Bin81}   K. Binder, Z.~Phys.~B -- Cond.~Mat. \textbf{43}, 119 (1981).
\bibitem{Bin81_2} K. Binder, Phys.~Rev.~Lett.\ \textbf{47}, 693 (1981).
\bibitem{Bin84}   K. Binder and D.~P. Landau, Phys.~Rev.~B \textbf{30}, 1477 (1984).
\bibitem{Vol93}   K. Vollmayr et al.\ Z.~Phys.~B \textbf{91}, 113 (1993).
\bibitem{Lan00}   D.~P. Landau and K. Binder, \textit{A guide to Monte~Carlo simulations in statistical physics} (Cambridge University Press, Cambridge, United Kingdom, 2000).
\bibitem{Mos02}   S. Mossa et al.\ Phys.~Rev.~Lett. \textbf{88}, 138701 (2002).
\bibitem{Cat05}   M. Catanzaro, M. Boguñá, and R. Pastor-Satorras, Phys.~Rev.~E \textbf{71}, 027103 (2005).
\end{thebibliography}
\end{document}